\documentclass[runningheads]{llncs}
\usepackage[T1]{fontenc}
\usepackage{graphicx}
\usepackage{url}
\usepackage{makecell}
\usepackage{multirow}
\usepackage{rotating}
\usepackage{threeparttable} 
\usepackage{amsmath}  
\usepackage{amsfonts,amssymb}

\usepackage[linesnumbered,ruled,vlined]{algorithm2e} 
\usepackage{algpseudocode}
\usepackage{algorithm2e}

\SetKwInput{KwInput}{Input}                
\SetKwInput{KwOutput}{Output}              

\usepackage{booktabs}
\usepackage{arydshln}
\usepackage[misc]{ifsym}
\usepackage{xcolor}
\newcommand{\jcx}[1]{\textcolor{black}{#1}}

\begin{document}
\title{EG-ConMix: An Intrusion Detection Method based on Graph Contrastive Learning}
%
%

\author{Lijin Wu\inst{1}\textsuperscript{(\Letter)}   \and
        Shanshan Lei\inst{1} \and
        Feilong Liao\inst{1} \and
        Yuanjun Zheng\inst{2} \and
        Yuxin Liu\inst{3} \and
        Wentao Fu\inst{4,5} \and
        Hao Song\inst{4,5} \and
        Jiajun Zhou\inst{4,5}
        }

\authorrunning{Wu et al.}
%
\institute{
Fujian Power Co., Ltd. Electric Power Research Institute, \\Fuzhou 350007, China
\and
State Grid Fujian Electric Power Co., Ltd. Fuzhou Power Supply Company,\\Fuzhou 350007, China
\and
State Grid Fujian Electric Power Co., Ltd. Putian Power Supply Company,\\Putian 351100, China
\and
Institute of Cyberspace Security, Zhejiang University of Technology, \\Hangzhou 310023, China  
\and
Binjiang Cyberspace Security Institute of ZJUT, \\Hangzhou 310056, China \\
\email{58794271@qq.com}}
%
\maketitle              
\begin{abstract}

\jcx{As the number of IoT devices increases, security concerns become more prominent. The impact of threats can be minimized by deploying Network Intrusion Detection System (NIDS) by monitoring network traffic, detecting and discovering intrusions, and issuing security alerts promptly. 
Most intrusion detection research in recent years has been directed towards the pair of traffic itself without considering the interrelationships among them, thus limiting the monitoring of complex IoT network attack events. Besides, anomalous traffic in real networks accounts for only a small fraction, which leads to a severe imbalance problem in the dataset that makes algorithmic learning and prediction extremely difficult. 
In this paper, we propose an EG-ConMix method based on E-GraphSAGE, incorporating a data augmentation module to fix the problem of data imbalance. 
In addition, we incorporate contrastive learning to discern the difference between normal and malicious traffic samples, facilitating the extraction of key features. 
Extensive experiments on two publicly available datasets demonstrate the superior intrusion detection performance of EG-ConMix compared to state-of-the-art methods. Remarkably, it exhibits significant advantages in terms of training speed and accuracy for large-scale graphs.}

\end{abstract}

\section{Introduction}
With the continuous development and popularization of the Internet of Things (IoT) technology, more and more devices and objects are connected to the Internet. 
According to statistics released by authoritative organizations~\cite{Statistics}, it is expected that more than 75 billion connected devices will be put into use by 2025. 
Nowadays, IoT technology is deeply integrated with artificial intelligence, big data, cloud computing, and other technologies. 
\jcx{This integration has significantly enhanced the convenience and efficiency of individuals' lives and work across diverse domains, including smart cities, healthcare, homes, transportation, and more.}

However, as the number of IoT devices continues to increase, IoT security issues are becoming increasingly prominent. 
As IoT devices are usually designed with low-cost hardware and software, they are less secure and prone to hacking and malware threats, leading to privacy leakage and data loss of users, and even causing serious harm to national security, life and property security, and other aspects.
In recent years, a series of IoT security incidents have emerged. 
In 2016, the U.S. Internet domain name resolution service provider (DYN) suffered a hack in which attackers used the Mirai virus to infect IoT devices and build a botnet to launch a distributed denial-of-service (DDoS) attack, leading to widespread service disruptions~\cite{XAQY201611009}. 
In 2019, the Microsoft Security Response Center disclosed APT28 malicious intrusion into IoT devices, where hackers used default passwords and vulnerabilities to cause leakage of IP information of targeted users~\cite{APT28}.
In 2022, the IoT system of Empresas Publicas de Medellin (EPM), Colombia's largest utility energy supplier, was hit by a black cat ransom attack~\cite{Blackcat}, which resulted in the disruption of a large number of networked devices and services and had a significant impact on many users. 
As a result, IoT security has become a topic of great interest to cope with the growing security needs and to start addressing the security threats faced by IoT devices in various scenarios.

In order to protect the security of IoT devices and users, timely detection and discovery of intrusion behavior and immediate adoption of corresponding defensive measures are important.
By deploying Network Intrusion Detection System (NIDS), different traffic in the network can be monitored, effective information can be efficiently extracted 
to find out whether there is an intrusion threat
and security alarms can be issued in time
to minimize the impact of the threat. There are two main types of NIDS: feature-based systems and anomaly detection-based systems. 
\jcx{Feature-based NIDS identify attacks by pattern matching in monitored network traffic using a predefined attack feature set. While proficient at recognizing known attacks, this NIDS type exhibits limitations in identifying novel attacks or modified versions of existing attack patterns.}
In contrast, anomaly detection-based systems, which identify attacks by detecting deviations in network traffic patterns, have the potential to detect novel attacks and can be subdivided into statistical analysis-based methods~\cite{lee1998data}, cluster analysis-based methods~\cite{khan2007new}, artificial neural network-based methods~\cite{hodo2016threat}, or deep learning-based methods~\cite{diro2018distributed}. 
\jcx{The advancement of machine learning technology, notably the extensive utilization of deep learning in recent years, enhances the self-learning, self-adaptive, and generalization capabilities of network intrusion detection systems. 
This progress enables more effective detection of unknown network attack behaviors.}


\jcx{Recently, Graph Neural Network (GNN) methods have been widely adopted in various fields to mine in-depth features, which leads to an increasing number of methods to model network traffic data as graphs. 
The network traffic commonly is conducted analyses in the form of flows. These flows are identified by communication endpoints, including IP addresses, port numbers, and transport protocols. Furthermore, they are annotated with a set of flow fields, such as packet byte counts and durations. Specifically, communication endpoints are mapped as graph nodes, and network traffic is mapped as edges of the graph, enriched with topological information. The information encapsulated in these edges plays a pivotal role in tasks related to network traffic classification and the detection of anomalous traffic patterns.
}

In applications of anomaly detection, real-world data is usually distributed in an unbalanced manner due to the rarity of anomalous events relative to normal events. In addition, data quality issues limited by high cost of label acquisition, information loss, redundancy, and errors can also cause serious difficulties for most algorithms that assume that the data is relatively equilibrium distributed~\cite{zhou2022data}. Mapping-based methods~\cite{zhou2020data,zhou2020m} for graph data enhancement are applied to solve the problem of data size limitation.

\jcx{Above all, despite the existence of numerous anomaly detection methods for network traffic, several challenges persist.}
\jcx{There are still few applications of GNN in the field of intrusion detection. In addition, traditional machine learning methods mainly classify intrusions based on the statistical features of network flows, and do not utilise the inherent structural and topological information in network flows for network intrusion detection. }
\jcx{In GNN methods, the poor model performance is due to the fact that the graph method relies on data, which suffers from data imbalance.}

\jcx{In this paper, to address the aforementioned issues, we introduce the EG-ConMix method. This technique incorporates two novel modules into the E-GraphSAGE~\cite{lo2022graphsage} framework, which are the data augmentation module and the contrast learning module. The data augmentation module utilizes the enhanced Mixup technique, called multi-pattern Mixup (MP-Mixup), to generate additional positive samples, alleviating the imbalance problem with the multi-pattern network traffic data. Furthermore, the augmented data is employed to develop specific contrastive learning methods that enhance the discrimination between positive and negative samples.}
\jcx{In this way, by introducing data augmentation and contrast learning modules, our framework achieves state-of-the-art performance in intrusion detection.}

The major contributions of our work are summarized as follows:
\begin{itemize}
  \item[$\bullet$] \jcx{We introduce the EG-ConMix algorithm, which explores the topological patterns of network intrusions in the Internet of Things (IoT) by modeling network traffic as a graph to enhance detection capabilities.}
  \item[$\bullet$] \jcx{For mitigating the data imbalance issue, We develop the Multi-Pattern Mixup (MP-Mixup) enhancement technique to create additional positive samples based on various patterns in the network traffic graph.}
  \item[$\bullet$] \jcx{Utilizing the augmented samples, we conduct contrast learning to derive feature representations capable of discerning distinct categories.}
  \item[$\bullet$] We applied the method to 2 benchmark IoT NIDS datasets for network intrusion detection, and the results demonstrate its potential through extensive experimental evaluation.
\end{itemize}
The rest of the paper is organized as follows. Section 2 discusses the key related work, deep learning based intrusion detection and related research on GNN based intrusion detection. Some of the preliminaries, construction of network traffic graphs and details of intrusion detection algorithms and models introducing Mixup and contrastive learning mechanisms are presented in Section 3. Section 4 gives the experimental analysis, including the introduction of the dataset, the experimental setup, and the analysis and discussion of the results. Section 5 summarizes the full paper and looks at future directions.

\section{Related Work}

\subsection{Classical Intrusion Detection Methods}
Thanks to the rapid development of neural networks, deep learning (DL) methods have become a key technology in various fields. Unlike traditional machine learning approaches, deep learning-based methods are able to identify important hidden features and representations from input data, avoiding feature engineering that requires the experience of domain experts. Deep learning makes the learning process more efficient and for intrusion behavior detection in complex networks, there are unsupervised methods based on multiple variants of Deep Belief Networks (DBN)~\cite{hinton2006fast} or Auto-Encoders (AE)~\cite{vincent2010stacked} or supervised methods based on Convolutional Neural Networks (CNN)~\cite{lecun1995convolutional}, or traditional Multi-Layer Perceptrons (MLP)~\cite{zhang2021deep}. 

While DL models have achieved significant success in learning new patterns, they are trained from inherently flat data structures (e.g., vectors or lattices) and are unable to capture the complex structural patterns necessary to detect Advanced Persistent Threats (APTs) and zero-day attacks. Indeed, such threats are often characterized by new attack patterns involving weak signals that disappear over an arbitrary period of time. Graphs provide a high-level and abstract generic representation; on the one hand, this graph-structured data can encode complex point-to-point relationships to learn richer informative representations; on the other hand, domain-specific knowledge incorporated in the structural and semantic information of the original data can capture finer-grained relationships between the data. Compared to flat data, graphs provide more semantics and release valuable relational information, which is ubiquitous in cyber-attack scenarios. Indeed, intrusions are essentially characterized by a series of suspicious and benign interactions between entities (e.g., hosts in a network or processes in a host system). When represented as graphs, these interactions can be learned by Graph Representation Learning (GRL) models, some of which have achieved excellent results in various cybersecurity tasks, such as vulnerability detection~\cite{nguyen2022regvd}, Ponzi detection~\cite{jin2022heterogeneous}, Blockchain decentralization~\cite{2022Behavior}, or malware detection~\cite{norouzian2021hybroid}.

In recent years, there has been a growing interest in developing graph-based deep learning algorithms, both unsupervised and supervised. Among these graph-based deep learning algorithms, GNN has been designed as a powerful deep graph representation learning technique for encoding non-Euclidean graph data to facilitate representation learning, downstream classification and prediction tasks. For example, GraphSAGE~\cite{hamilton2017inductive} implements inductive learning and small batch training through neighborhood sampling. Unlike methods that directly compute classification results, the core of GraphSAGE is to learn the mapping function that generates the embedding vectors of nodes, which extends GCN into an inductive learning task with greater scalability and generalization to unknown nodes, so that when a new node appears, it is only necessary to compute the embedding vectors of the node through the already learned mapping function.

\subsection{Self-supervised Learning
} 
In self-supervised learning, researchers have explored how to design and build more appropriate data enhancement processes. Mixup~\cite{zhang2017mixup} is a regularization technique and a data augmentation method proposed by Zhang et al to improve the generalization ability of neural networks. The idea is to randomly select two sample pairs $(x_{i},y_{i})$ and $(x_{j},y_{j})$ in the dataset $D$, obtain a convex combination $(\hat{x},\hat{y})$ of the samples and labels through Eq 6, where $y_{i}$ and $y_{j}$ are the unique heat codes of the corresponding labels, and subsequently train the network on the convex combination of the samples.
 
Compared to supervised learning, contrastive learning is a self-supervised learning method that utilizes the properties of the data itself to obtain supervised information and explores the intrinsic connections of the data for network training. The core idea of learning general features of a dataset by having the model learn which data points are similar or different is that two augmented views of the same data are treated as positively sampled to bring them closer together, while all other instances are treated as negatively sampled to make them farther apart. 
In recent years, self-supervised contrastive learning has achieved great success in graph learning tasks. DGI~\cite{velivckovic2018deep} aims to learn node representations by deriving patch representations and corresponding high-level graph summaries using an established graph convolutional network architecture and maximizing the mutual information between the two. Anomaly detection methods based on self-supervised contrast learning, such as CoLA~\cite{liu2021anomaly}, have made good progress. This method not only alleviates the model's dependence on labeled data, but also effectively weakens the effect of noise on detection and reduces computational complexity by learning the local connection information in the attribute network, i.e., constructing pairs of positive and negative instances between a node and its neighboring substructures (subgraphs), and avoiding direct input of the entire network into the model.


\section{Methodology}

\subsection{Network Traffic Graphs Construction}

\begin{figure}[htp]
\centering
  \includegraphics[width=\linewidth]{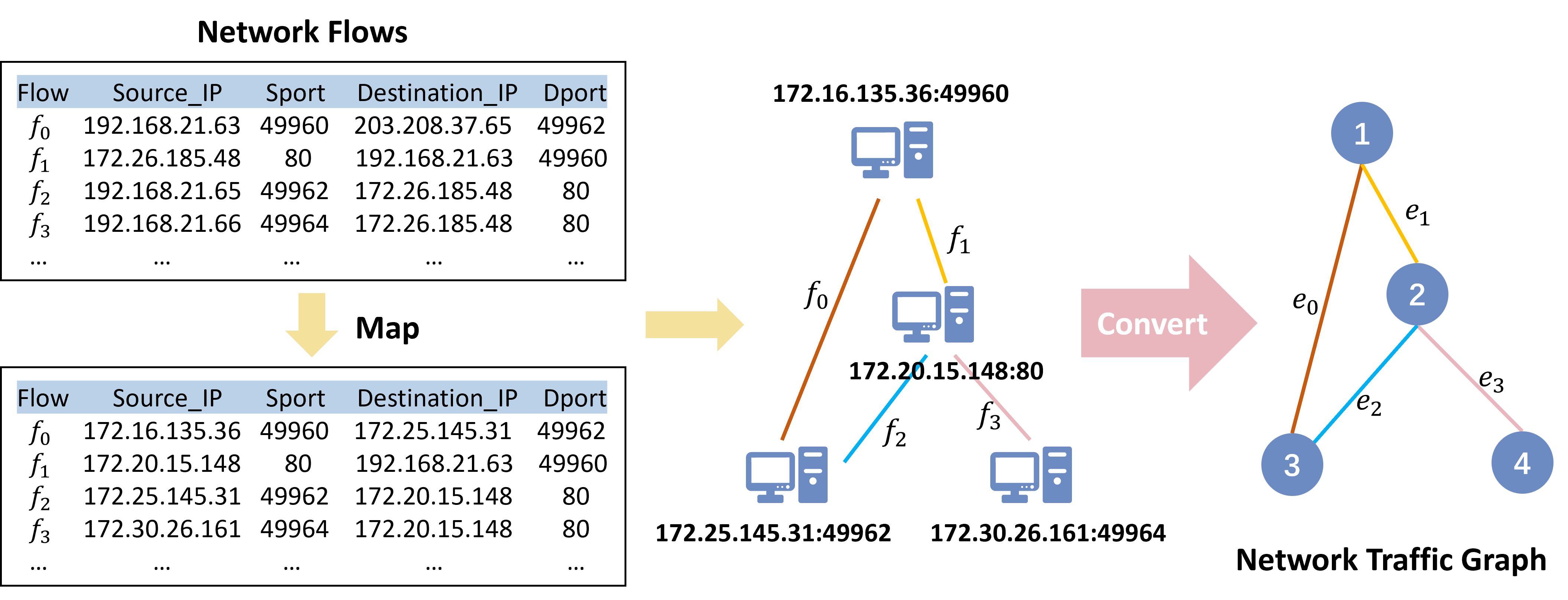}
  \caption{By mapping the source and destination IP addresses in the raw data to a defined range. The IP address and port number are bound as a 2-tuple to represent the node, and the traffic transmitted between the two devices is represented as an edge. In this way, the network traffic is transformed into a graph structure representation.}
  \label{fig:network traffic graph}
\end{figure}

Network traffic graphs are constructed to convert network traffic data into graph structures for intrusion detection using graph neural networks (GNN). Network traffic data typically contains multiple fields that identify the source and destination of the communication as well as provide detailed information about the traffic such as the number of packets, number of bytes, duration, and so on. This process first requires identifying the endpoints of the network traffic graph, which in this paper is defined using the following four flow fields: source IP address, source (L4) port, destination IP address, and destination (L4) port. Each of these four fields forms two 2-tuples that are used to identify the source and destination nodes. That is, source and destination IP addresses and port numbers, the network devices represented by these endpoints are mapped as nodes in the graph. The corresponding traffic transmitted between the source and destination nodes can be mapped as edges of the network traffic graph, such as the data exchanged between the source node (172.25.145.31: 49962) and the destination node (172.20.15.148: 80). The process of constructing a network traffic graph is shown in Fig.~\ref{fig:network traffic graph}.

To construct the graph, the original source IP addresses are randomly mapped to a predefined range of IP addresses (e.g., from 172.16.0.1 to 172.31.0.1), mitigating the potential issue of limited IP address diversity in NIDS datasets. Direct use of raw IP addresses may inadvertently introduce bias into the model training process by associating specific IP addresses with attack traffic. Consequently, the model could become overly sensitive to traffic originating from these addresses, leading to misclassification of normal traffic as malicious. By employing random mapping of original source IP addresses to a designated IP address range, we prevent the model from learning specific IP address associations. This randomization enhances the model's ability to extract more generalized features, thereby maintaining robust performance against novel attacks. Furthermore, this mapping strategy safeguards data privacy by obscuring sensitive information inherent in the original network traffic.

In graph construction, except for the four key flow fields used to define edges, all other flow record fields are assigned to edges, such as information about the number of packets, number of bytes, and duration. Nodes, on the other hand, are given all-1 vectors, indicating that they have no features. This treatment allows the model to focus on extracting information from edges (flows) rather than node features.

In this way, through the above steps, the network traffic data is converted into a graph structure where nodes represent devices in the network and edges represent communication traffic between devices. The graph contains not only the topology of the network traffic, but also the specific characteristics of the traffic, and this graph representation provides a natural data structure and rich information for the subsequent E-GraphSAGE model to perform the intrusion detection task effectively.

\subsection{Model Architecture}
\begin{figure}[htp]
	\centering
  \includegraphics[width=\textwidth]{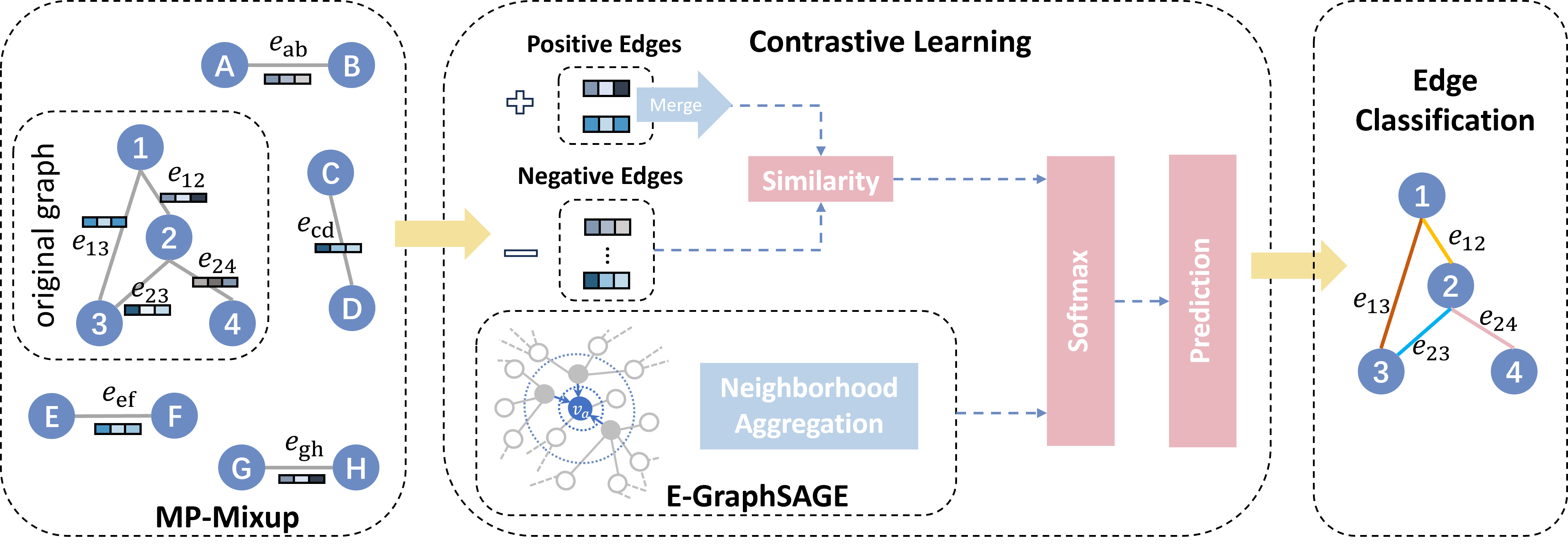}
  \caption{The architecture of EG-ConMix. The complete workflow proceeds as follows:
           1) For the data imbalance problem, MP-Mixup is used to generate multiple pairs of virtual nodes and corresponding connected edges;
           2) Minimize the distance of positive sample pairs and maximize the distance of negative sample pairs by comparing the similarity of positive and negative samples;
           3) The results of contrastive learning are combined with the E-GraphSAGE model to complete the classification of edges for the purpose of intrusion detection.
           }
  \label{fig:framework}
\end{figure}

The framework of the network intrusion detection model proposed in this paper is shown in Fig.~\ref{fig:framework}. The model is mainly divided into three parts: (1) MP-Mixup data augmentation module; (2) contrastive learning-based sample representation module; and (3) E-GraphSAGE-based anomalous traffic detection and classification module.

\subsubsection{Multi-Pattern Mixup}
Due to the extreme imbalance of the dataset, which leads to poor model learning performance, this paper uses the Mixup method which is used to perform data augmentation on the data to alleviate the data imbalance problem. However, considering that the learning method of the model in this paper is different from the traditional Mixup method, the Multi-pattern Mixup method used in this paper improves the original Mixup method. The traditional Mixup method is suitable for node-level tasks, which augment the data by generating new nodes. The model in this paper uses an edge-level approach and therefore requires the use of Multi-pattern Mixup to sample edges with a positive-to-negative ratio of $\sigma$ for data augmentation. In Multi-pattern Mixup method, the final enhancement of the generated edges connects the nodes with the presence of two types of edges, i.e., harmful edges - harmful edges and harmful edges - unharmful edges. Since the objective of this paper is a binary classification task, Mixup is improved in terms of labeling to handle the two types of edges generated. In this paper, an indicator function is introduced to define the labels of the edges between harmful nodes to unharmful nodes by judging the $\lambda$ value, when the $\lambda$ value is greater than 0.5 (when it is less than 0.5), the unharmful edge is greater than (less than) the harmful edge, and the indicator function determines the label to be 0 (1), i.e., the harmless edge (harmful edge). The formula is as follows:

\begin{equation}
\begin{cases}\hat{x} = \lambda x_{i}+(1-\lambda)x_{j}\\\hat{y} = \mathbb{I}(0<\lambda<0.5)
\end{cases}
\end{equation}

where $x_{i}, x_{j}$ denote unharmful and harmful edges, respectively, and $(\hat{x},\hat{y})$ denotes the convex combination of samples and labels, $\lambda\sim \operatorname{Beta}(\alpha,\alpha),\alpha\in(0,\infty),\lambda\in[0,1].$



\subsubsection{EG-ConMix}
In order to be able to better utilize the large amount of data generated by Mixup augmentation, this paper uses the contrastive learning method to enhance the learning effect of the model. Contrastive learning is a self-supervised learning method that utilizes the properties of the data itself to obtain supervised information and explore the intrinsic connections of the data for network training. In this paper, we use the InfoNCE loss~\cite{2014Notes} method to extract the virtual positive samples obtained after mixup enhancement, combine them with the positive samples in the dataset to generate new pairs of positive samples, and calculate the similarity between the positive samples and the negative samples randomly sampled from the dataset to achieve the purpose of comparative learning, he ratio of positive and negative samples in this process is $\gamma$. InfoNCE loss formula is shown as follows:

\begin{equation}
\mathcal{L}_{k}= -\frac{1}{N}\sum_{e^{+}\in{\mathcal{E^{+}}}}\log(\frac{\exp(e^{+}\cdot e_{uv})}{\exp(e^{+}\cdot e_{uv})+\sum_{e^{-}\in{\mathcal{E}}}\exp(e^{-}\cdot e_{uv})})  
\end{equation}

Where $e_{uv}$ is a positive sample edge in the set of edges $\mathcal{E}, \mathcal{E^{+}}$ is the set of positive samples in the set of edges $\mathcal{E}$, $e^{+}\in{\mathcal{E^{+}}}$, and $\mathcal{E^{-}}$ is the set of negative samples in the set of edges $\mathcal{E}$, and $e^{-}\in{\mathcal{E^{-}}}$.

The GNN approach primarily focuses on propagating node features through message passing, but it lacks the capability to classify edges using edge features. The E-GraphSAGE algorithm~\cite{lo2022graphsage}, addresses this limitation by extracting k-hop edge features and generating edge embeddings, thus enabling effective traffic classification for network intrusion detection.
The method proposed in this paper is further improved based on GraphSAGE. The main difference with the original GraphSAGE algorithm lies in the three parts of algorithm input, message propagation function and algorithm output. The input of EG-ConMix includes edge features for edge feature message propagation, and the contrastive learning method is also introduced to enhance the model learning effect.
The model in this paper performs neighbor information aggregation based on edge features rather than node features, and the neighbor aggregator function used, as shown in Eq 3.

\begin{equation}
\mathbf{h}_{\mathcal{N}(\nu)}^k\leftarrow\mathrm{~AGG}_k\left(\left\{\mathbf{h}_u^{k-1}\ \| \ \mathbf{e}_{u\nu}^{k-1},\forall u\in\mathcal{N}(\nu),u\nu\in\mathcal{E}\right\}\right)
\end{equation}

Where $\mathbf{e}_{u\nu}^{k-1}$ denotes the edge feature ${u\nu}$ from $N(v)$, the sampled neighborhood  of node $v$, at layer $k-1$. The aggregated edge features can be concatenated with neighborhood node embeddings $\mathbf{h}_u^{k-1}$ for edge feature message propagation. In line 5, the node embeddings of node $v$ at layer $k$ are computed based on the k-hop edge features, and the neighboring node embeddings consist of neighboring edge features. As a result, the k-hop edge features and topological patterns in the graph are collected and aggregated and can be combined with the node representation of the current node $\mathbf{h}_v^{k-1}$.

The final node representation at depth $K$ is computed, $\mathbf{z}_{\nu}=\mathbf{h}_{\nu}^{K}$.The final embedding ${z}_{u\nu}^K$ of each edge is computed as a concatenation of pairs of node embeddings $u$ and $v$, as shown in Eq 4.
\begin{equation}
\mathbf{z}_{u\nu}^K\leftarrow\mathrm{~CONCAT}{\left(\mathbf{z}_u^K,\mathbf{z}_\nu^K\right)},u\nu\in\mathcal{E}
\end{equation}

\subsubsection{Model Train} During training, the EG-ConMix model optimizes the model parameters by minimizing the loss function (cross entropy loss):

\begin{equation}
  \mathcal{L}_{c}=\frac{1}{N}\sum_i-[y_i\cdot \log(p_i)+(1-y_i)\cdot \log(1-p_i)],
\end{equation} 
where $y_{i}$ denotes the label of sample $i$, with positive samples being 0, and negative samples being 1. $p_{i}$ denotes the probability of sample $i$ predicting a positive sample.

The model loss function includes a contrast loss function and a cross-entropy loss function, which are expressed as follows:
\begin{equation}
  \mathcal{L} =  \mathcal{L}_\textit{c} \ + \theta\mathcal{L}_\textit{k} ,
\end{equation}

\section{Experiments}
In order to evaluate the performance of the proposed model in this paper, we conduct intrusion detection experiments on two IoT datasets. First, the datasets used in the experiments are briefly described. Then, we discuss the detailed experimental setup, including metrics and implementation. Subsequently, we conduct ablation experiments on the proposed approach and analyze and discuss the impact of Mixup and contrastive learning on the experimental results.

\subsection{Datasets}
To verify that our proposed method is effective for network intrusion detection, we conducted experiments on two publicly available real-world IoT datasets, and the specifications of the two IoT datasets are shown in Table~\ref{tb:data}.


\begin{table}[h!]
  \renewcommand\arraystretch{1.2}
  \centering
  \caption{Dataset properties.}
  \label{tb:data}
  \resizebox{0.9\linewidth}{!}{
  \fontsize{7pt}{10pt}\selectfont
  \begin{tabular}{ccccc}
\hline\hline
Dataset & Size    & Nodes     & Edges     & ~~Features~~  \\ 
\hline
BoT-IoT & 2.33GB  & 3,683,084 & 3,668,522 & 55        \\
~~ToN-IoT~~ & ~~575.4MB~~ & ~~1,005,577~~ & ~~1,155,447~~ & 30        \\
\hline\hline
\end{tabular}}
\end{table}

\begin{itemize}
  \item[$\bullet$] \textbf{BoT-IoT~\cite{koroniotis2019towards}:} \  Created by Koroniotis et al. in 2019, specifically for IoT networks. Features were extracted by modeling IoT services (e.g., water stations) using the Node-red tool and extracting features using the Argus tool. The dataset consists of 6 types of attacks and a total of 47 features labeled with the corresponding categories. The dataset contains only 477 (0.01\%) benign flows and 3,668,045 (99.99\%) attack flows, totaling 3,668,522 flows.
  
  \item[$\bullet$] \textbf{ToN-IoT~\cite{alsaedi2020ton_iot}:} \ Created by Abdullah et al. in 2019, ToN-IoT contains different types of IoT data such as operating system logs, telemetry data from IoT/Industrial IoT services, and IoT network traffic collected from a network at the University of New South Wales in Canberra, Australia. Only the network traffic portion of the dataset is used in this paper. The dataset consists of 796,380 (3.56\%) benign flows and 21,542,641 (96.44\%) attack flows, totaling 22,339,021 flows.
\end{itemize}

\subsection{Comparison Methods}
To examine the performance of the model, we compared it with three commonly used anomaly detection methods. In order to examine the performance of the model, we compare it with three commonly used anomaly detection methods. 
\begin{itemize}
\item[$\bullet$] \textbf{E-GraphSAGE:} \
The original E-GraphSAGE algorithm without the introduction of data augmentation and contrast learning.
\item[$\bullet$] \textbf{E-GraphSAGE-Res~\cite{chang2021graph}:} \
E-GraphSAGE-Res deals with the category imbalance problem by adding residual connections to preserve the original information and improve the detection performance of a few categories, which is based on E-GraphSAGE. 
\end{itemize}

\subsection{Experimental Settings}
\subsubsection{Data Preparation}
We divide each dataset with the same proportion for training, validation as well as testing, and the proportions of the training set, validation set, and testing set are \{70\%, 10\%, 20\%\}, respectively. For positive and negative samples in the training set, validation set, and test set, we use a fixed proportion for sampling to ensure the same proportion of positive and negative samples in the divided dataset. We repeat the 10-fold cross-validation five times and report the F1-macro and its standard deviation to ensure that the results obtained by our model are credible under the unbalanced dataset. It is worth noting that in each evaluation we used the same test data for each method to allow for fair comparisons.

\subsubsection{Evaluation Metrics}
The overall accuracy performance of the model is measured by the F1 score, a metric that calculates the reconciled mean of sensitivity and accuracy, and is a widely used evaluation metric for models trained using unbalanced datasets, denoted as follows:
\begin{equation}
  F1-score=2\times\frac{Recall\times Precision}{Recall+Precision},
\end{equation}
Where precision measures the ability of an intrusion detection system to identify only attacks, recall can be thought of as the ability of the system to discover all attacks. The higher the F1 score, the better the balance between precision and recall achieved by the model. In order to clearly distinguish the differences and benefits of our model in anomaly detection, we compared the macro F1 scores between several different algorithms. When calculating macro F1 scores, the F1 scores of all categories are averaged. This way, we do not discriminate categories based on their percentages and increase the importance of a few categories. The macro F1 score is used as a key performance metric for our experiments because this metric addresses potential imbalances within the dataset. In addition, the standard deviation was used for further performance evaluation.

\subsubsection{Parameter Configuration}
The EG-ConMix algorithm proposed in this paper is implemented based on PyTorch and its corresponding PyTorch Geometric library. The algorithm contains two layers of E-GraphSAGE layer, in which we use 128-dimensional hidden units corresponding to the dimensions of node embedding. Meanwhile, in order to optimize the neural network and prevent the overfitting problem, we added a dropout mechanism between the two E-GraphSAGE layers and set its parameter to 0.2. We also used the ReLU activation function as the nonlinear transformation function of the model and the Softmax function as our classifier. For optimizer and loss function selection, we used the Adam optimizer, set its learning rate to 0.01 to perform gradient descent in the backpropagation stage, and chose cross entropy and InfoNCE loss as our loss function. Other than that, other hyperparameters are given in Table~\ref{tb:Parameter Configuration}.


\begin{table}[h!]
  \renewcommand\arraystretch{1.2}
  \centering
  \caption{Parameter Configuration.}
  \label{tb:Parameter Configuration}
  \resizebox{1.0\linewidth}{!}{
    \begin{tabular}{ccc} 
\hline\hline
Hyperparamter           & Descrption                                                        & default  \\ 
\hline
$\alpha$                   & Mixup positive and negative inter-sample weight coefficients      & 0.3      \\
$\beta$                    & Mixup positive inter-sample weight coefficients                   & 0.2      \\
$\gamma$ & Sample size of positive and negative samples in contrast learning & 10       \\
$\sigma$            & The number of positive and negative samples sampled in mixup      & 200       \\
\hline\hline
\end{tabular}
}
\end{table}

\subsection{Evaluation on Intrusion Detection}
The proposed model is experimentally compared with the baseline method with the same dataset to evaluate its intrusion detection performance. The experimental results are shown in Table~\ref{tb:f1-result}.

\begin{table}
\renewcommand\arraystretch{1.2}
\centering
\caption{Performance on intrusion detection task reported in macro-f1 and standard deviation.}
\label{tb:f1-result}
 \resizebox{0.9\textwidth}{!}{%
\begin{tabular}{ccccc} 
\hline\hline
\multirow{2}{*}{Method} & \multicolumn{2}{c}{BoT-IoT}           & \multicolumn{2}{c}{ToN-IoT}  \\ 
\cline{2-5}
                        & \multicolumn{1}{l}{~~Macro-F1~~} & Std    & ~~Macro-F1~~        & Std        \\ 
\hline
E-GraphSAGE             & 0.9515                       & ~~0.0094~~ & 0.9008          & ~~0.0015~~     \\
~~E-GraphSAGE-Res~~         & 0.9515                       & 0.0094 & 0.9121          & 0.0237     \\
EG-Con         & 0.9515                       & 0.0094 & 0.9184          & 0.0113     \\
EG-Mix         & 0.9521                       & 0.0101 & 0.9165          & 0.0136     \\
EG-ConMix      & \textbf{0.9562}              & 0.0133 & \textbf{0.9268} & 0.0010     \\
\hline\hline
\end{tabular}
}
\end{table}

The table lists the evaluation results for the intrusion detection task, from which it can be seen that our EG-ConMix achieves state-of-the-art results compared to previous methods. Specifically, our EG-ConMix significantly outperforms the E-GraphSAGE-based correlation algorithm baseline on all datasets, which suggests that the edge vector representations obtained after Mixup data augmentation and contrastive learning do play an important role in identifying anomalous edges. Our EG-ConMix outperforms the strong baseline: on both datasets, we observe macro F1 scores of 95.62\% and 92.68\%, respectively.

\begin{figure}[t]
	\centering
  \includegraphics[width=\textwidth]{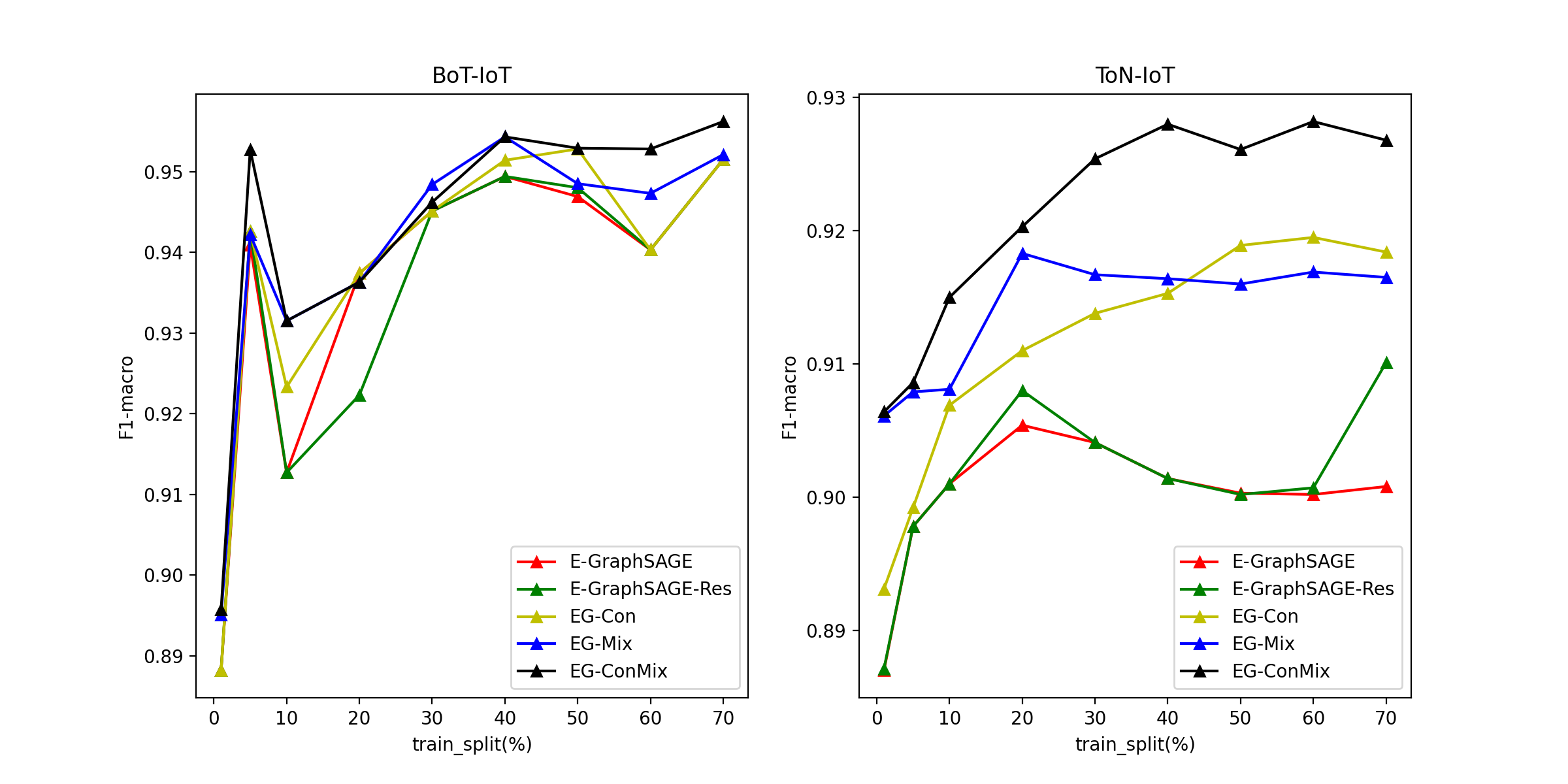}
  \caption{ Statistics of macro-f1 score results for data partitioning experiments based on multiple methods.
  }
  \label{fig:BoT-ToN}
\end{figure}

To further demonstrate the superior performance of the EG-ConMix model over other benchmark models, we set up dataset partitioning experiments to validate the effectiveness of our contrastive learning and data augmentation experiments. Fig.~\ref{fig:BoT-ToN} shows the macro F1-score and its standard deviation results obtained by our model when the training set is divided into \{1\%, 5\%, 10\%, 20\%, 30\%, 40\%, 50\%, 60\%, 70\%\}. As can be seen from the figure, EG-ConMix and its corresponding EG-Con and EG-Mix methods outperform the benchmark model on both datasets. In the BoT-IoT dataset, we observe that there are fluctuations in the results obtained by the model as the training set keeps growing. However, in fact, the fluctuation is also present in the benchmark model and our model results are more stable than the benchmark model, so this result also proves the effectiveness of the contrastive learning and Mixup methods in the face of extremely unbalanced data.

\subsection{Parametric Analysis}
\begin{figure}[t]
	\centering
  \includegraphics[width=\textwidth]{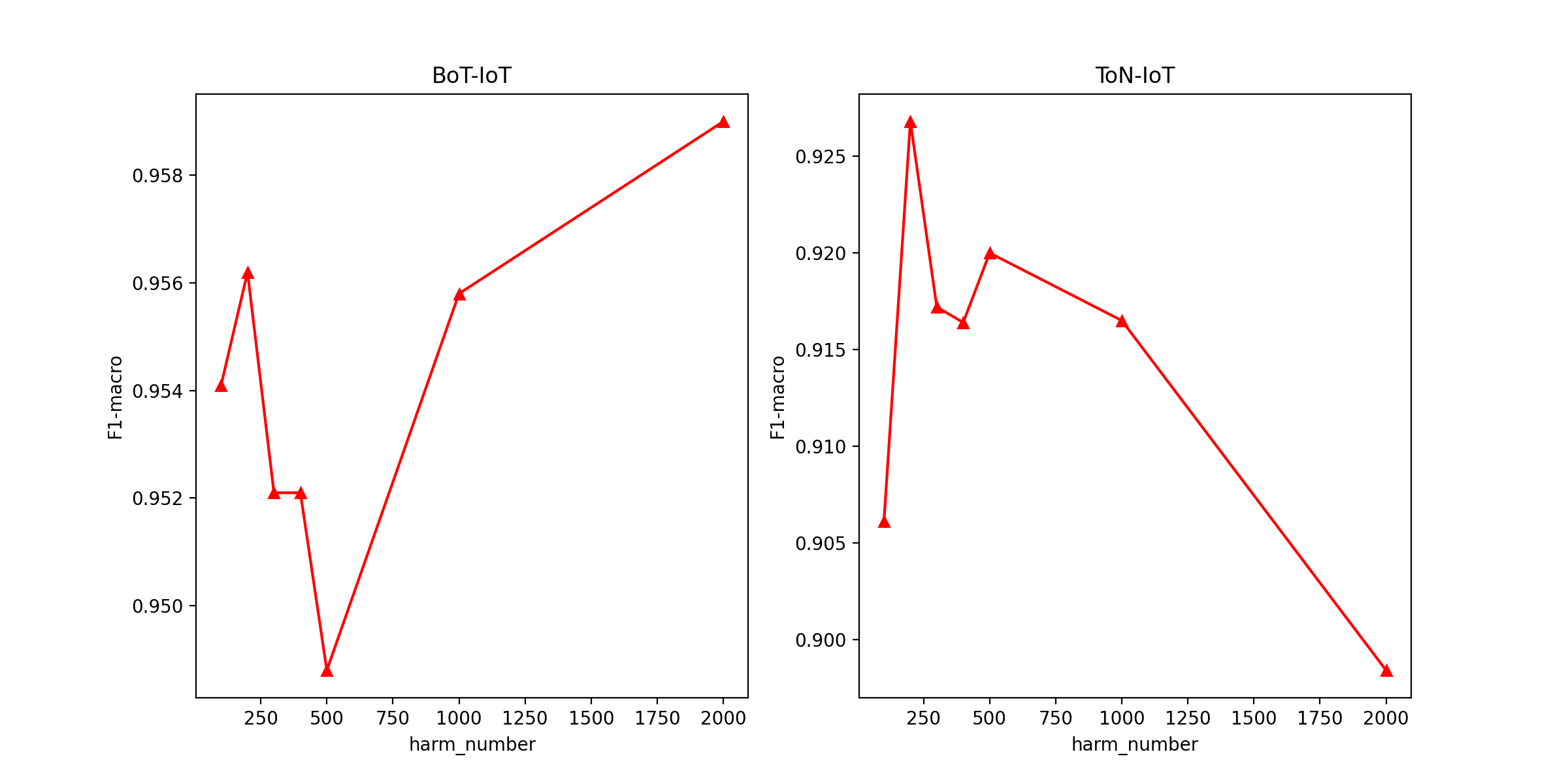}
  \caption{Analysis of the number of negative samples corresponding to positive samples of Mixup data augmentation.
  }
  \label{fig:Parameter}
\end{figure}

In order to ensure that the method we use has a better performance, we conducted parametric experiments on the number of samples of Mixup, which is divided into \{100, 200, 300, 400, 500, 1000, 2000\}, and the results are shown in Fig.~\ref{fig:Parameter}. We observe that in the BoT-IoT dataset, there are fluctuations in the results of the model with the increase in the number of samples, and therefore, we hypothesize that Mixup's data augmentation method is not optimal for the data augmentation method for this extremely imbalanced dataset. As for the ToN-IoT dataset, the model's value reaches its highest when the number of samples reaches 200 and keeps decreasing thereafter, which we believe is due to the fact that the ratio of positive and negative samples in this dataset is more balanced compared to the BoT-IoT dataset, and that large-scale data augmentation for it will eventually lead to its sample reverse imbalance. Although, the model has higher performance when the number of samples in the BoT-IoT dataset reaches 2000, large-scale data augmentation leads to a reduction in computational efficiency. Therefore, considering the model effect and computational efficiency, we use 200 as the sampling parameter of our Mixup method.

\section{Conclusion}
In intrusion detection applications for IoT networks, detection methods with the help of graph neural networks are relatively less studied and understood. In this paper, we improve the E-GraphSAGE algorithm and in turn propose the EG-ConMix algorithm. To address the problem of data imbalance in the real world, we propose to augment the network traffic data by Mixup method, which solves the data imbalance and introduces contrastive learning, which compares the similarity between the samples to learn the representation of the samples, which helps detecting anomalies and recognizing the threats, and thus improves the network security. Meanwhile, our experiments on two publicly available datasets demonstrate that our framework can achieve intrusion detection on the web with satisfactory predictive performance on benchmark datasets, with outstanding advantages in training speed and accuracy on large-scale datasets.


\bibliographystyle{splncs04_}
\bibliography{sample-base-ori}

\end{document}